# Mott gapping in *3d* $ABO_3$ perovskites without Mott-Hubbard interelectronic U

Julien Varignon[1], Manuel Bibes[1] and Alex Zunger[2]

[1]*Unité Mixte de Physique, CNRS, Thales, Université Paris Sud, Université Paris-Saclay, 91767, France*

[2]*Energy Institute, University of Colorado Boulder Colorado 80309, Boulder, CO, USA*

**Abstract**

The existence of band gaps in Mott insulators such as perovskite oxides with partially filled *3d* shells has been traditionally explained in terms of strong, dynamic inter-electronic repulsion codified by the on-site repulsion energy U in the Hubbard Hamiltonian. The success of the “DFT+U approach” where an empirical on-site potential term U is added to the exchange-and correlation Density Functional Theory (DFT) raised questions on whether U in DFT+U represents interelectronic correlation in the same way as it does in the Hubbard Hamiltonian, and if empiricism in selecting U can be avoided. Here we illustrate that *ab-initio* DFT *without any U* is able to predict gapping trends and structural symmetry breaking (octahedra rotations, Jahn-Teller modes, bond disproportionation) for all $ABO_3$ *3d* perovskites from titanates to nickelates in both spin-ordered and spin disordered paramagnetic phases. We describe the paramagnetic phases as a supercell where individual sites can have different local environments thereby allowing DFT to develop finite moments on different sites as long as the total cell has zero moment. We use a recently developed exchange and correlation functional (“SCAN”) that is sanctioned by the usual single-determinant, mean-field DFT paradigm with static correlations, but has a more precise rendering of self-interaction cancelation. Our results suggest that strong dynamic electronic correlations are not playing a universal role in gapping of *3d* $ABO_3$ Mott insulators, and opens the way for future applications of DFT for studying a plethora of complexity effects that depend on the existence of gaps, such as doping, defects, and band alignment in $ABO_3$ oxides.

## I. Introduction

Transition metal oxide perovskites $ABO_3$ with a *3d* transition metal atom substituting the B site exhibit intriguing metal *vs* insulator characteristics as well as different forms of magnetism across the series both in their high- temperature (HT) spin-disordered paramagnetic (PM) phases and/or in the low-temperature (LT) spin-ordered phases[1]. Some of the compounds are metallic ($CaVO_3$, $SrVO_3$ or $LaNiO_3$ for instance), while others are insulating, such as titanates $RTiO_3$ ($d^1$), vanadates $RVO_3$ ($d^2$), manganites $CaMnO_3$ ($d^3$) and $RMnO_3$ ($d^4$), ferrites $CaFeO_3$ ($d^4$) and $RFeO_3$ ($d^5$), cobaltites $RCoO_3$ ($d^6$), nickelates $RNiO_3$ ($d^7$) or possibly cuprates $RCuO_3$ ($d^8$), where R is a rare-earth element or yttrium. Concomitantly with the opening of a band gap, one observes a variety of systematic symmetry-breaking modes, such as the Jahn-Teller (JT) distortions in $RVO_3$[2] or $RMnO_3$[3] compounds, propagating either in-phase ($Q_2^+$ mode) or in anti-phase ($Q_2^-$ mode), or the B-O bond disproportionation $B_{oc}$ observed in the insulating phase of $RNiO_3$[4] and $CaFeO_3$[5] (see sketches in Figure 1). Understanding and controlling such band gaps and the associated lattice distortions and forms of magnetism is central to the ability to dope these oxides, just as is designing specific band offsets in oxide heterojunctions, to the benefit of future oxide electronics. The crucial question here is what minimum theoretical framework is needed to explain and therefore design such gapping-related phenomena.

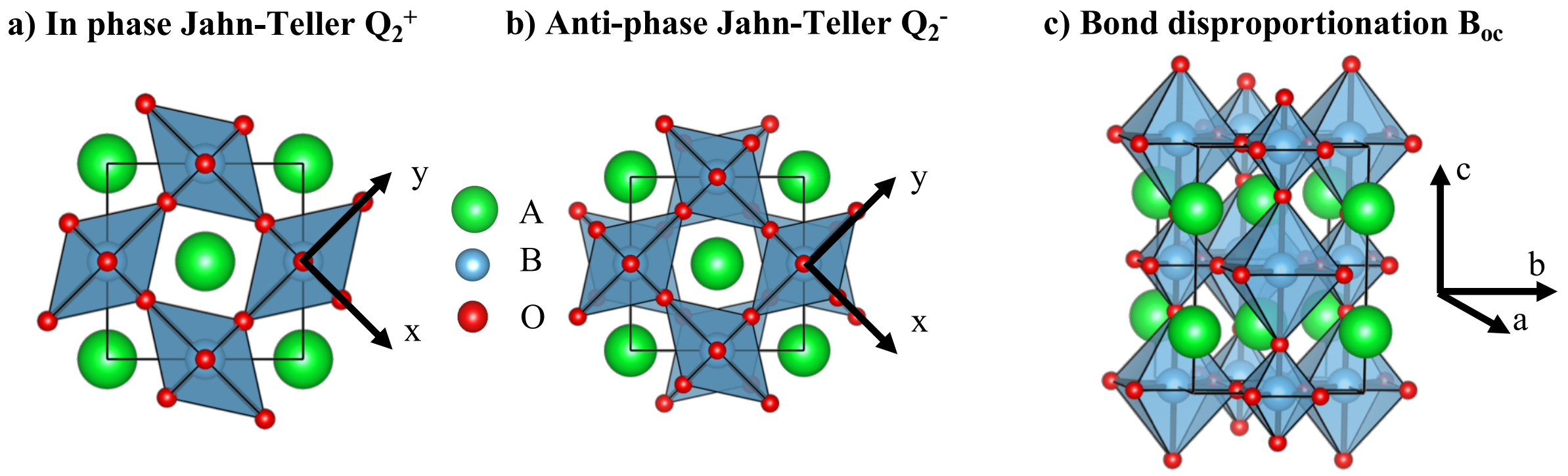

*Figure 1: Sketches of Jahn-Teller motions (a and b) and bond disproportionation (c) distortions appearing in some $ABO_3$ materials.*

The standard explanation of gapping in these compounds despite the presence of partially filled *d* shells and the ensuing expected orbital degeneracy is generally formulated in terms of strong inter-electronic repulsions appearing in the celebrated Mott-Hubbard model[6–9]. This leads to a uniform explanation of gapping for all *d*-electron $ABO_3$ compounds and degrees of spin order or disorder, based on the symmetry-conserving interelectronic repulsion, and ensuing localization. Within this framework, the experimental observations of a variety of different symmetry-breaking modes, such as those presented in Fig.1, or magnetic moments is not related to the gapping mechanism but can appear afterwards as an additional effect.

Whereas Density Functional Theory (DFT) has been shown to be able to address numerous physical effects in such oxides, including ferroelectricity[10], catalysis[11] and electrical battery voltage [12], its use of a single Slater determinant and its mean-field treatment of electron-electron interactions (static correlations) has, according to numerous literature statements[13–15], disqualified it for the study such "strongly correlated" oxides, requiring far more computationally costly dynamically correlated methodologies, such as Dynamical Mean Field Theory (DMFT).

However, the DFT calculations in such demonstrations of failure [13–15] often used a non-spin polarized description and at times exchange correlation functionals that do not distinguish occupied from unoccupied orbitals (LDA or GGA functionals without U), and generally neglect sublattice displacements. For example, Ref.[14] demonstrated vanishing band gaps in $LuNiO_3$, in contradiction with experiment, and Ref.[15] demonstrated failure to stabilize JT distortions in $LaMnO_3$, again, in contradiction with both experiments and DMFT calculations. Such naïve (N-) DFT calculations, however, do not necessarily represent what proper DFT can do, as demonstrated in recent calculations for the binary *3d* oxides MnO, NiO, CoO and FeO[16] or $ABO_3$ materials[17–24]. In these calculations [16,24], the PM phase, which is a collection of magnetic moments $\vec{m}$ with random magnitude and direction on each site i but whose sum is zero ($\vec{M} = \sum_i \vec{m}_i = \vec{0}$), was represented by specially constructed supercells that has a total zero spin, but unlike the N-DFT implementations, there was no requirement that each transition metal ion have a zero spin. Such polymorphous representation lowered substantially the total energy relative to the N-DFT representations, while producing finite band gaps and local moments in all studied binary and ternary $ABO_3$ compounds, in general accord with experiment. The need for an exchange-correlation functional that distinguishes occupied from unoccupied orbitals was satisfied by using a simplified self-interaction corrected functional in the form of "DFT+U" [25]. The use of "U" created sometimes the false impression that this approach owes its success in explaining gaps to the interelectronic repulsion and localization, just as in the Mott-Hubbard view.

To examine if the explanation of gapping requires an explicit Hubbard U, we have performed DFT calculations on several popular $ABO_3$ perovskite oxides with *d* fillings from 1 to 8 electrons, using (unlike previous DFT+U calculations of e.g. Ref.[24]) the recently developed SCAN meta-GGA functional[26], *without Mott-Hubbard interelectronic* U. The use of the SCAN functional for transition metal and transition metal oxides is reported in many papers[27–32]. Unlike these publications we do not just study how SCAN functional behaves with ternary oxides on gaps, structural features or electron localization, but we tackle a different problem: is the Mott-Hubbard model required to capture trends in gapping, structural motions or magnetic moments in $ABO_3$ materials? Here we show that (i) DFT, without any U parameter but with an exchange-correlation functional better representing self-interaction errors, is sufficient to explain trends in properties of ternary $ABO_3$ oxides in both LT spin-ordered and HT PM phases; (ii) since DFT-no-U and its static mean-field treatment of electron interactions are largely sufficient to produce insulation, $ABO_3$ oxide perovskites may certainly be complicated but they are not necessarily strongly dynamically correlated and (iii) thus, gapping in such specific cases may not sweepingly obey the celebrated Mott-Hubbard explanation of formation of two electron sites which depends on the existence of d like band edges and on interelectronic repulsion exceeding band width, neither of which are needed in the current explanation . However, our results do not imply that if DFT works for a material, then all forms of correlations do not play a role.

## II. **Method**

We have performed DFT[33,34] total energy minimization with respect to lattice parameters and cell-internal atomic positions of different perovskite oxides with *d* fillings from 1 to 8 electrons. Structure types compared for their total energy included orthorhombic (*Pbnm*), monoclinic ($P2_1/b$ and $P2_1/n$) and rhombohedral (*R-3c*), whereas spin configurations examined included ferromagnetic (FM), as well as classical A, C and G-type antiferromagnetic (AFM) orders, and more complex S-type AFM order based on ↑↑↓↓ spins chains in the (*ab*)-plane with different stackings along the c axis. We emphasize here that we did not explore the relative stability of the different magnetic orders and we just focused on the spin order experimentally observed at low temperature for each compound. The paramagnetic (PM) spin-disordered state has been modelled using the Special Quasirandom Structures (SQS) method[35], following references[16,24], which represents a random "alloy" of up and down (collinear) spins with total spin zero, within the 2x2x2 orthorhombic or monoclinic cells (32 $ABO_3$ formula units containing 160 atoms). Convergence with SQS supercell size was tested and found adequate at 160 atoms[24]. All atomic displacements as well as breaking of degeneracies of partially occupied $e_g$ and $t_{2g}$ levels are allowed as long as they reduce the total energy. Amplitudes of the energy-minimizing distortions were determined by performing a symmetry adapted mode analysis [36,37] with a reference structure set to the ideal high symmetry cubic *Pm-3m* structure of perovskites. The lattice parameter of this hypothetical cubic structure is fixed to the ground state pseudo-cubic lattice parameter.

## III. ***Results***

***Calculated structural and electronic properties:*** Figure 2 summarizes the calculated band gap ΔE (in eV), magnetic moment $M_{3d}$ (in $\mu_B$) associated with the B cations and amplitudes of distortions (in Å) associated with JT and bond disproportionation motions, all done without Hubbard U, compared with experimental values available in literature. The calculated energy-minimizing lattice type agrees with experiments with the exception of $YVO_3$ and $LaVO_3$ in the PM phase and $CaFeO_3$ in the AFM phase. In the two vanadates, the strongly entangled spin and orbital degrees of freedom induce small lattice distortions on each octahedra in the PM phase (where each transition metal element experiences a unique potential), thus reducing the symmetry from *P2₁/c* to *P-1*[38]. However, the lattice parameters, B-O-B angles and B-O bond lengths are very similar to the respective quantities observed in experimental structures. For $CaFeO_3$, the AFM-S magnetic order that we have used to approximate the experimentally observed AFM spiral[5] breaks the inversion center and induce some small lattice distortions such as polar displacements[39], thus producing a polar $P2_1$ space group instead of a centrosymmetric $P2_1/n$ symmetry.

| | *d* filling | Sym. | Mag. | $\Delta E_{NM}$ (mev/f.u) | Electronic properties: $E_g$ (eV) | Electronic properties: $M_{3d}$ ($\mu_B$) | Structural distortions: $Q_2^+$ (Å) | Structural distortions: $Q_2^-$ (Å) | Structural distortions: $B_{oc}$ (Å) |
|---|---|---|---|---|---|---|---|---|---|
| $YTiO_3$ | $t_{2g}^1 e_g^0$ | *Pbnm* | FM | -242 | 0.08 | 0.92 (0.84[a]) | 0.03 | - | - |
| | | *Pbnm* | PM | -218 | 0.33 (1.20[b]) | 0.84 | 0.02 (0.00[c]) | - | - |
| $LaTiO_3$ | $t_{2g}^1 e_g^0$ | *Pbnm* | AFMG | -75 | 0.05 | 0.67 (0.46[d]) | 0.06 | - | - |
| | | *Pbnm* | PM | -84 | 0.14 (0.20[b]) | 0.78 | 0.04 (0.04[c]) | - | - |
| $YVO_3$ | $t_{2g}^2 e_g^0$ | *Pbnm* | AFMG | -1052 | 0.89 | 1.77 (1.72[e]) | 0.19 (0.14[e]) | - | - |
| | | *P-1* | PM | -978 | 0.55 (1.60[f]) | 1.82 | NA | NA | - |
| $LaVO_3$ | $t_{2g}^2 e_g^0$ | *Pbnm* | AFMC | -864 | 0.78 | 1.78 (1.30[g]) | 0.00 (0.01[h]) | 0.09 (0.08[h]) | - |
| | | *P-1* | PM | -833 | 0.36 (1.10[i]) | 1.80 | NA | NA | - |
| $CaMnO_3$ | $t_{2g}^3 e_g^0$ | *Pbnm* | AFMG | -2049 | 0.79 | 2.62 (2.64[j]) | 0.01 | - | - |
| | | *Pbnm* | PM | -2036 | 1.46 (ins.) | 2.62 | 0.01 (0.04[k]) | - | - |
| $LaMnO_3$ | $t_{2g}^3 e_g^1$ | *Pbnm* | AFMA | -1783 | 0.52 | 3.65 (3.70[l]) | 0.28 | - | - |
| | | *Pbnm* | PM | -1758 | 0.30 (0.24[m]-1.70[n]) | 3.60 | 0.32 (0.30[o]) | - | - |
| $CaFeO_3$ | $t_{2g}^3 e_g^0 + t_{2g}^3 e_g^2$ | *$P2_1$* | AFMS | -1474 | 0.20 | 2.72-3.67 (2.48-3.48[p]) | 0.00 | - | 0.13 |
| | | *$P2_1/n$* | PM | -1449 | 0.07 (0.25[q]) | 2.66-3.72 | 0.00 (0.04[p]) | - | 0.14 (0.18[p]) |
| $LaFeO_3$ | $t_{2g}^3 e_g^2$ | *Pbnm* | AFMG | -1073 | 1.67 (2.10[i]) | 3.94 (4.60[r]) | 0.01 | - | - |
| | | *Pbnm* | PM | -936 | 0.52 | 4.04 | 0.01 (0.00[s]) | - | - |
| $YCoO_3$ | $t_{2g}^6 e_g^0$ | *Pbnm* | NM | - | 1.48 (ins.[t]) | - | 0.06 (0.05[t]) | - | - |
| $YNiO_3$ | $t_{2g}^6 e_g^2 + t_{2g}^6 e_g^0$ | *$P2_1/n$* | AFMS | -353 | 0.92 | 1.41-0.00 (1.70-0.40[u]) | 0.05 | - | 0.18 |
| | | *$P2_1/n$* | PM | -350 | 0.59 (0.20[i]-1.00[v]) | 1.38-0.28 | 0.05 (0.05[w]) | - | 0.17 (0.13[w]) |
| $LaCuO_3$ | $t_{2g}^6 e_g^2$ | *R-3c* | AFMG | -143 | 0.48 (?) | 0.70 (?) | - | - | - |

***Figure 2: Key properties of oxide perovskites with the SCAN meta-GGA functional without U using SQS for the PM phase (Experimental values are provided in parentheses)*** *Energy difference (in meV/f.u) between spin polarized and N-DFT (NM) solution, band gap $E_g$ (in eV), magnetic moment $M_{3d}$ (in $\mu_B$) associated with the B cation and amplitudes of distortions (in Å) associated with Jahn-Teller and bond disproportionation motions* [34]. *Ins. stands for insulating phases. Experimental values are taken from a: Ref.*[40]*, b: Ref.*[41]*, c: Ref.*[42]*, d: Ref.*[43]*, e: Ref.*[44]*, f: Ref.*[45]*; g: Ref.*[46]*, h: Ref.*[47]*, i: Ref.*[48]*, j: Ref.*[49]*, k: Ref.*[50]*, l: Ref.*[51]*, m: Ref.*[52]*, n: Ref.*[53]*, o: Ref.*[54]*, p: Ref.*[5]*, q: Ref.*[55]*, r: Ref.*[56]*, s: Ref.*[57]*, t: Ref.*[58]*, u: Ref.*[4]*, v: Ref.*[59]*, w: Ref.*[60]

***Electronic structure as gleaned from DOS:*** Figure 3 depicts the B-atom (green) and Oxygen (red) projected density of states (in eV/states/f.u) averaged on all sites in the low temperature phase. Several observations are apparent: (i) only the perovskites with light B atoms ($LaTiO_3$ and $YVO_3$, Fig3.a and b) have band edge states with dominant *d* characters (as imagined in the original Mott-Hubbard model), while the O *p* levels lie below the B *d* levels in accord with Pavarini *et al* [61,62]; (ii) for heavier *3d* atoms the *d* bands move deeper in energy, towards the oxygen bands and the electronic structure becomes strongly *p-d* hybridized, yielding covalent charge transfer insulator behaviors ($CaMnO_3$ to $YNiO_3$ in accord with Bisogni *et al* [63]). Thus, our calculation reproduces the correct position of transition metal *d* levels with respect to O *p* levels as deduced from experiments[46,64] and DMFT simulations on some perovskite compounds [1,61–63].

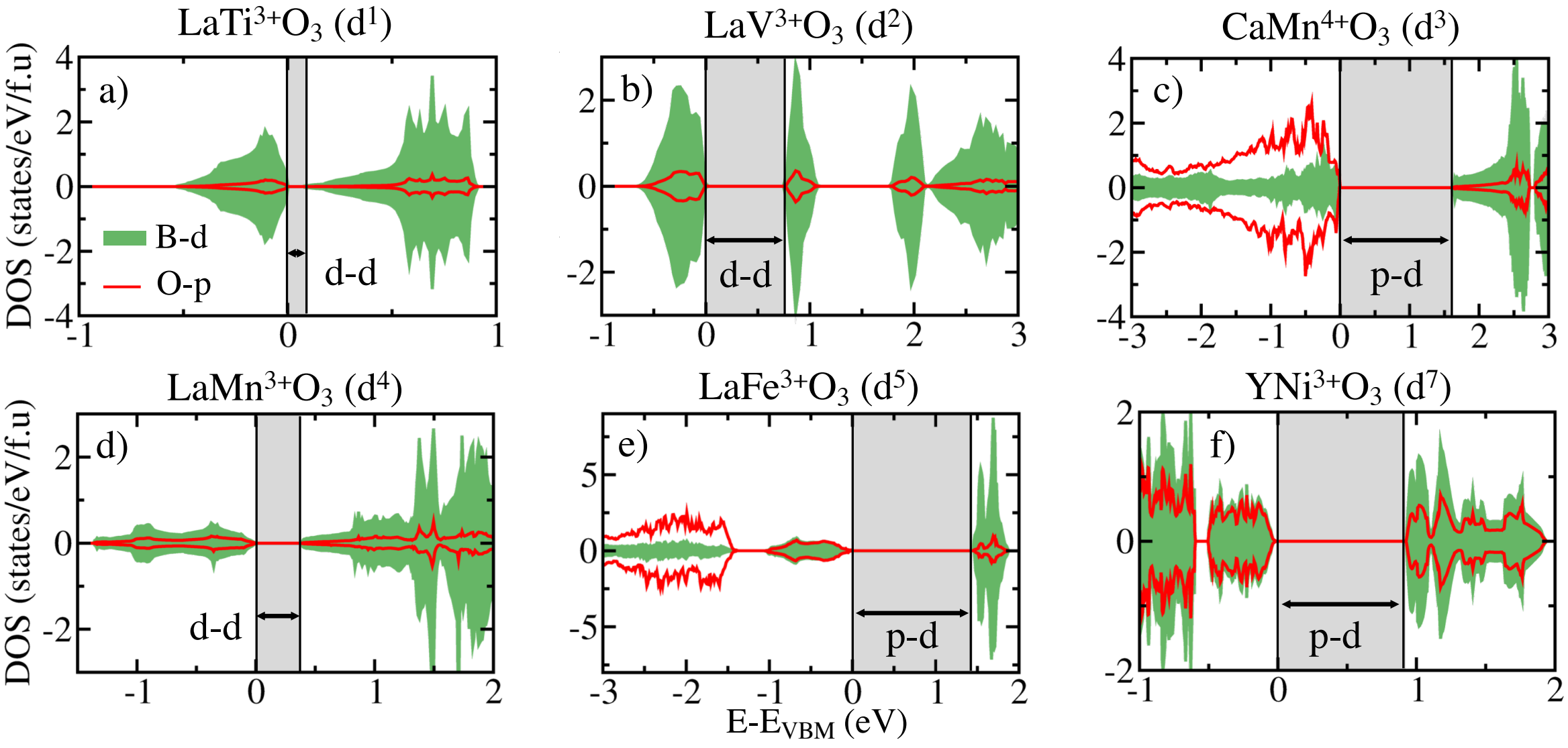

*__Figure 3:__ Projected density of states (in states/eV/f. u) on B-d levels (filled green) and O-p levels (red line) averaged on all sites in the low temperature phase. Band edges are shown with the vertical lines. The band gap is represented by the light grey area.*

***Band gaps without U:*** All $ABO_3$ compounds tested here are insulators in both their spin-ordered and spin-disordered PM solutions (Figure 2). These results agree with the insulating character observed experimentally (some experimental values available in literature are reported in Figure 2) and also reproduce trends observed with DFT+U and DMFT simulations on some of these materials (*e.g.* $RTiO_3$, $RVO_3$, $RMnO_3$ and $RNiO_3$, see references therein). For instance, we observe insulation in the yttrium nickelate $YNiO_3$ compound in both AFM and PM phases, as DMFT does in $LuNiO_3$ PM phase[14,65]. Likewise, we also observe an increase in the band gap when going from rare-earth titanates ($d^1$) to rare-earth vanadates ($d^2$) in agreement with experimental observations[41,45]. We emphasize here that experimental data of structural and electronic properties on bulk and stoichiometric $LaCuO_3$ crystals are scarce and diverging hindering confirmation of the SCAN-DFT calculation. Finally, just as standard exchange and correlation functionals underestimate the band gap of the highly uncorrelated semi-conductors such as Si and GaAs, the SCAN functional behaves similarly for $ABO_3$ materials and one may improve the band gap description and related quantities by using GW corrections [66].

***Magnetic moments:*** Our calculations provide magnetic moment values that are comparable to experimental quantities available in literature (see Figure 2). For instance, we capture the decrease of the magnetic moment of Ti cations when going from $YTiO_3$ to $LaTiO_3$[40,43]. Due to the presence of a DLE for Ni and Fe cations in $YNiO_3$ and $CaFeO_3$, respectively, two very different magnetic moments are extracted from our simulations. These quantities, compatible with experimental values, point towards a disproportionation of the unstable 3+ and 4+ formal oxidation states (FOS) of Ni and Fe cations, respectively, towards their more stable 2+/4+ and 3+/5+ FOS.

***Trends in energy differences between different phases:*** The energy gain in forming local moments is given by the total energy difference $E_{NM}$ -$E_{AFM}$. As the number of unpaired 3d electrons increases, we see that this energy strongly increases, signaling the large energy gain obtained by forming local magnetic moments, and thus the irrelevance of the NM ansatz (Fig.2). The energy cost for forming a random configuration from an ordered one is ($E_{PM}$-$E_{AFM}$)-TS where the first term is the contribution of the internal T=0 energy. For all compounds studied ($E_{PM}$-$E_{AFM}$)>0, *i.e.* the spin-ordered states are just slightly more stable than the PM solution. The entropy contribution will cause an order -disorder transition at finite $T_{Néel}$. $LaTiO_3$ is an exception in which ($E_{PM}$-$E_{AFM}$)<0, *i.e.* the experimentally observed AFM-G state is higher in energy than the PM state. This delicate balance could be because other magnetic configurations occur at low T. We checked the FM spin order (in the spirit of $YTiO_3$) for this compound and found that the FM order now represents an energy gain of 10 meV/f.u over the PM phase. Interestingly $LaTiO_3$ in the FM spin order is still an insulator ($E_g$=0.02 eV) with similar distortions to the PM and AFM-G solutions. It is possible that an exhaustive search for other spin configurations will change the result somewhat. Indeed, we did not perform such an exhaustive search.

***Electronic Localization without U:*** It is interesting to analyze the electronic localization of the *d* electrons in those perovskite oxides that are believed to be strongly dynamically correlated systems ($d^1$, $d^2$, $d^4$ and $d^7$ materials for instance). We report in Figure 4 the electronic charge density at the top of the valence band in the low temperature spin-ordered phase of $LaTiO_3$ ($d^1$), $YVO_3$ ($d^2$), $LaMnO_3$ ($d^4$) and $YNiO_3$ ($d^7$). In agreement with the pDOS reported in Fig.3, one notices increasing density on the O atoms when going from light to heavier transition metal atoms, underlining the progressive shift from “Mott insulators” ($LaTiO_3$, $YVO_3$, Fig4. a and b) to “charge transfer insulator” ($YNiO_3$, Fig4.d) behaviors as mentioned earlier. In $LaTiO_3$ (Fig4.a), the single $Ti^{3+}$ *d* electron is localized in a linear combination of the three $t_{2g}$ levels, whose relative coefficients are alternating on neighboring Ti sites. In $YVO_3$ (Fig4.b), the additional *d* electron sits in a combination of the $d_{xy}$ and $d_{xz}$ *or* $d_{yz}$ orbital, with alternating coefficients on neighboring sites due to the presence of an in-phase $Q_2^+$ JT motions in the low-temperature phase (see Figure 2). We observe a similar situation in $LaMnO_3$ (Fig4.c): the $Mn^{3+}$ “$e_g$” electron is localized either in $d_x^2$ or $d_y^2$ orbitals between neighboring Mn sites in the (*ab*)-plane but with similar stackings along the *c* axis. Finally, a “charge ordered” picture is observed in $YNiO_3$ (Fig4.d) with Ni cations sitting in an extended (compressed) octahedra bearing approximately 2 (0) electrons on the $e_g$ levels. This is clearly proving a charge disproportionated insulating state with Ni ions adopting their more stable 2+ and 4+ formal oxidation state (FOS) instead of the unstable 3+ FOS. Similar observations were obtained for other members such as $YTiO_3$, $LaVO_3$ or $CaFeO_3$, independently of the spin-order/disorder.

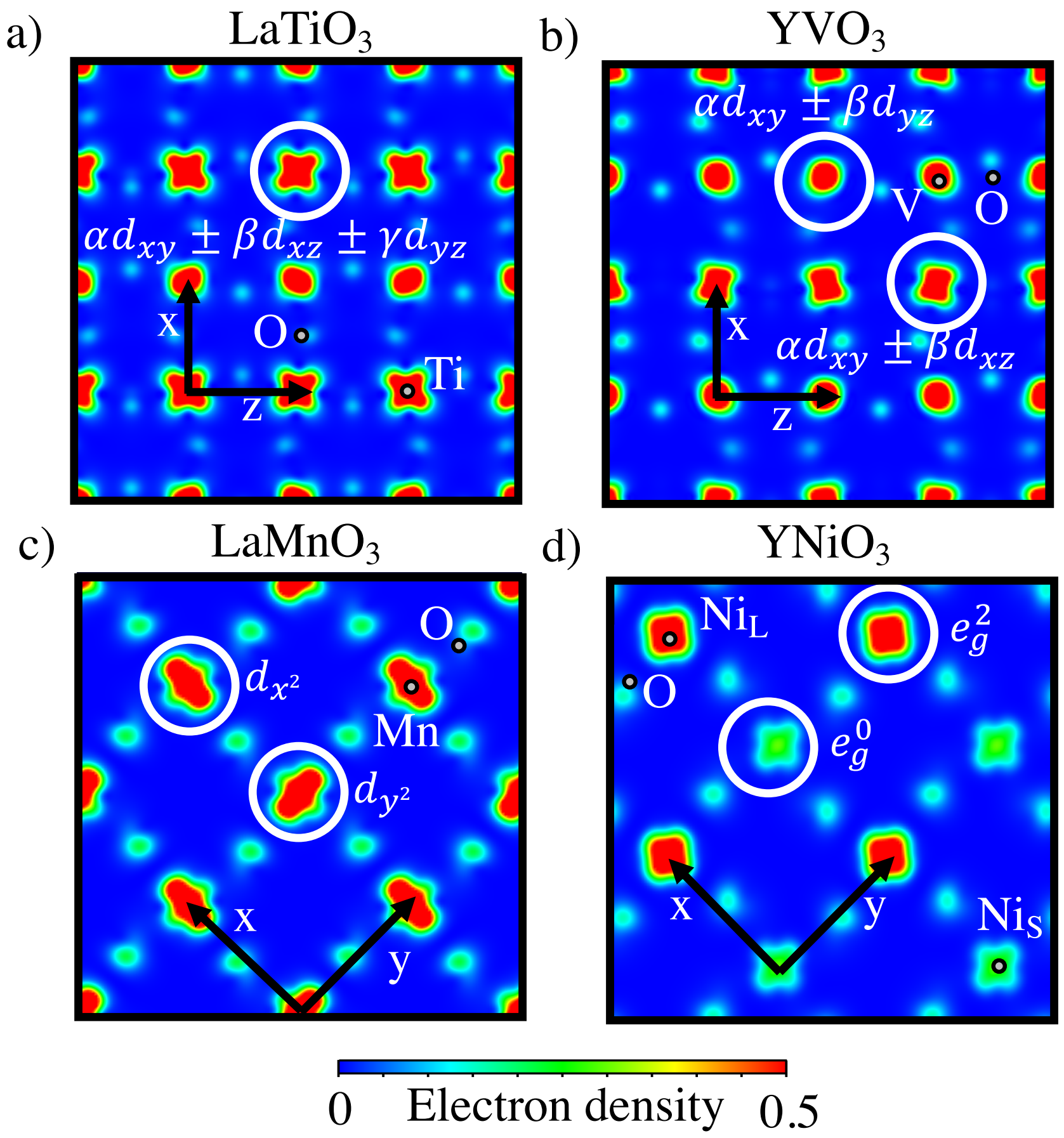

*Figure 4: Partial charge density maps of levels at the top of the valence band in some selected "correlated oxide perovskites".*

***How displacements affect or create gapping:*** To understand the role of JT and bond disproportionation motions on the gap opening, we use as a starting configuration a high symmetry *Pm-3m* phase and then apply to it successively all displacement modes appearing in the ground state AFM structure (*i.e.* $O_6$ rotations, antipolar displacements of ions). We then freeze the JT or $B_{oc}$ modes and compute the potential energy surface *vs* amplitude of the modes for NM, AFM and PM solutions in $LaMnO_3$ ($Q_2^+$ mode), $LaVO_3$ ($Q_2^-$ mode) and $YNiO_3$ ($B_{oc}$ mode). This protocol has been used in Ref. [15] to demonstrate the inability of naïve N-DFT to stabilize JT motions in $LaMnO_3$ and the crucial role of strong dynamic correlation to obtain such displacements. Our results for the same NM ansatz are reported in Figure 5 where we confirm that this naïve model yields a single well potential whose energy minimum is located in zero amplitude, *i.e.* these structural distortions do not appear. Comparing with the full DFT calculation in Fig.5 clearly shows, however that N-DFT is not what DFT can actually do, the latter producing distortions where they appear experimentally. Moving to AFM and PM spin-polarized solutions, we observe that the energy minimum of the different potentials is located to non-zero amplitude of JT and $B_{oc}$ modes. We conclude that DFT without an interelectronic U can stabilize these previously believed to be "correlation-induced lattice distortions"[14,15] if minimal ingredients are provided to the simulations. It is evident that dynamic correlation effects are not forcing these distortions whose role was previously hampered by a false initial hypothesis.

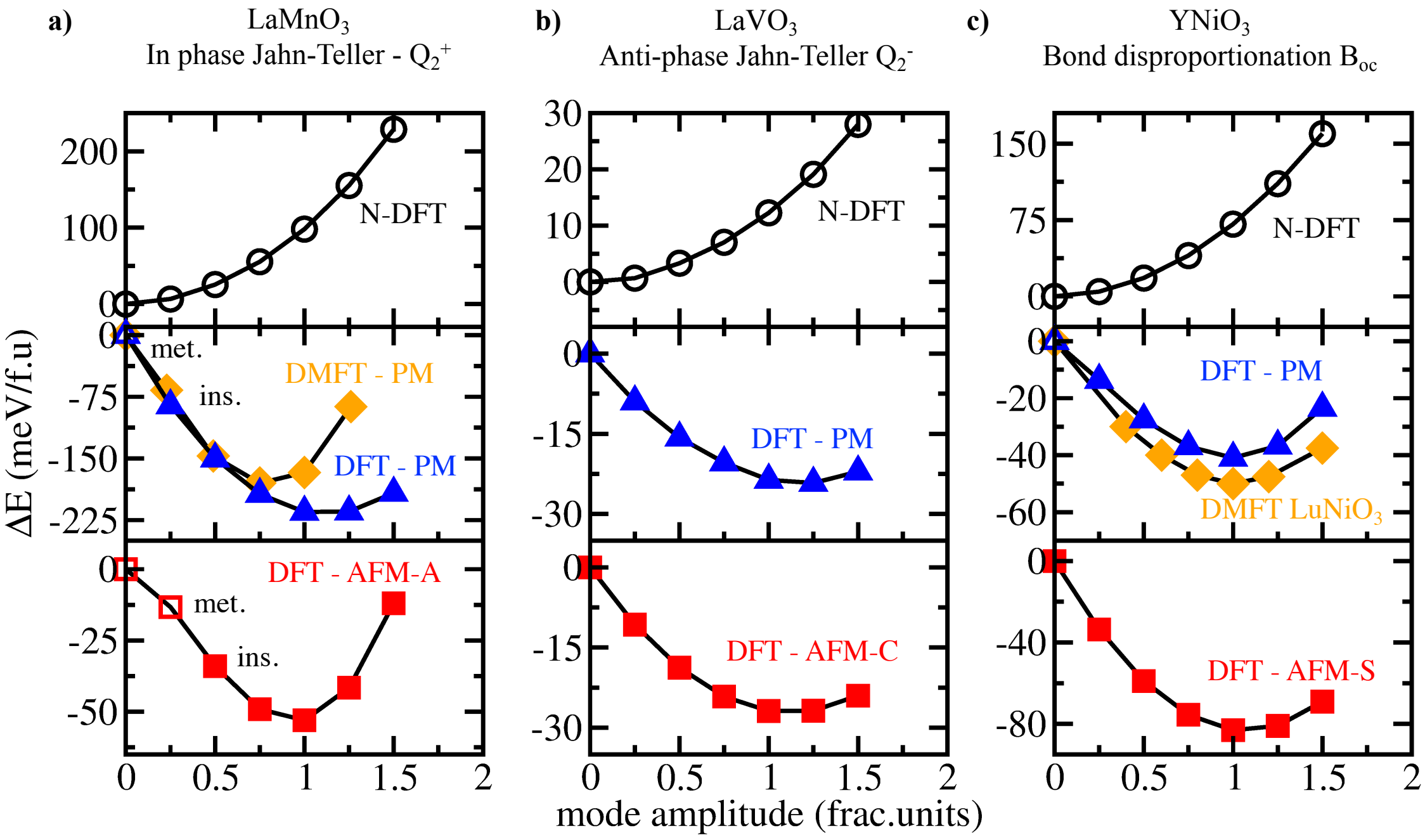

***Figure 5:*** *Energy gains ΔE (in meV/f.u with respect to the 0-mode amplitude) associated with Jahn-Teller distortions (a and b) and the bond disproportionation (c) modes as obtained by the naïve nonmagnetic N-DFT (black circles, upper panel), the PM described with a SQS supercell (blue triangles, middle panel) and AFM (red squares, lower panel). DMFT potential (orange diamonds) for the PM phase of $LaMnO_3$ and $LuNiO_3$ extrapolated from Refs.* [15,65] *are also reported. Filled and unfilled symbols correspond to insulating (ins.) and metallic (met.) solutions. The reference structure at 0 amplitude of JT or $B_{oc}$ modes is set to a material displaying octahedra rotations and anti-polar motions of ions.*

***The role of atomic displacements on the gap opening***. Opened and closed symbols in Fig.5 denote metallic or insulating solutions, respectively. Upon increasing the amplitude of the $Q_2^+$ and $B_{oc}$ modes in $LaMnO_3$ and $YNiO_3$, respectively, a band gap opens in the PM and AFM solutions. $LaVO_3$ is different where rotations plus antipolar displacements of ions are sufficient to produce an insulating state since the $Q_2^-$ JT mode is not important for the gap opening. The present SCAN-no-U results are consistent with DFT+U and DMFT simulations in $LaMnO_3$[13,23,67] and $RVO_3$[18,62], as well as the experimental observation of insulating states in $RVO_3$ irrespective of the presence of the JT $Q_2^-$ mode[68]. Surprisingly, even without any amplitude of the disproportionation $B_{oc}$ mode, $YNiO_3$ already exhibits a clear-cut of the Ni electronic structures with one Ni site bearing a magnetic moment of 1.06 (1.26) $\mu_B$ and the other one a value around 0.72 (0.00) $\mu_B$ in the PM (AFM) solution (Fig.5.c). It follows that $YNiO_3$ has a spontaneous tendency to undergo disproportionation effects through an electronic instability transforming the unstable 3+ formal oxidation state (FOS) of Ni cations to the more stable 2+ and 4+ FOS in the insulating phase. Nevertheless, only the AFM order can render an insulating state without bond disproportionation $B_{oc}$ mode (Fig.5.c) – this is consistent with the fact that the PM phase of $YNiO_3$ without bond disproportionation is found metallic in experiments[4]. The observation of an electronic instability agrees with our recent DFT+U study [24] and with numerous DMFT calculations identifying a spontaneous tendency of $Ni^{3+}$ cations to undergo disproportionation effects[65,69].

## IV. **Discussion**

So far, we have shown the ability of the SCAN functional to capture the formation of basic physical quantities (*e.g.* band gap, magnetic moments, structural motions) of $ABO_3$ materials upon various d fillings. We may now question the ability of this functional to reproduce physical trends within series showing isoelectronic configurations.

***The dependence of the Jahn-Teller distortions on the existence of the local magnetic moments: The case of $SrVO_3$*:** For instance, $SrVO_3$ ($d^1$) display a paramagnetic metallic state at all temperature while the isolectronic rare-earth titanates $RTiO_3$ exhibit insulating states. We have explored the paramagnetic state of $SrVO_3$ using the constrained magnetic approach proposed by Franchini *et al* [70,71] (we restricted $V^{4+}$ spins to be aligned along z, since spins have the tendency to flip during the self-consistent field). Unlike the isoelectronic $RTiO_3$ compounds, $SrVO_3$ relaxes to a cubic metallic phase – thus showing no JT distortion. This can be explained using the 1926 Goldschmidt tolerance factor close to 1, which implies cubic stability in agreement with experiments. In contrast, we have shown that isovalent $YTiO_3$ and $LaTiO_3$ are insulating for both paramagnetic and spin-ordered phases.

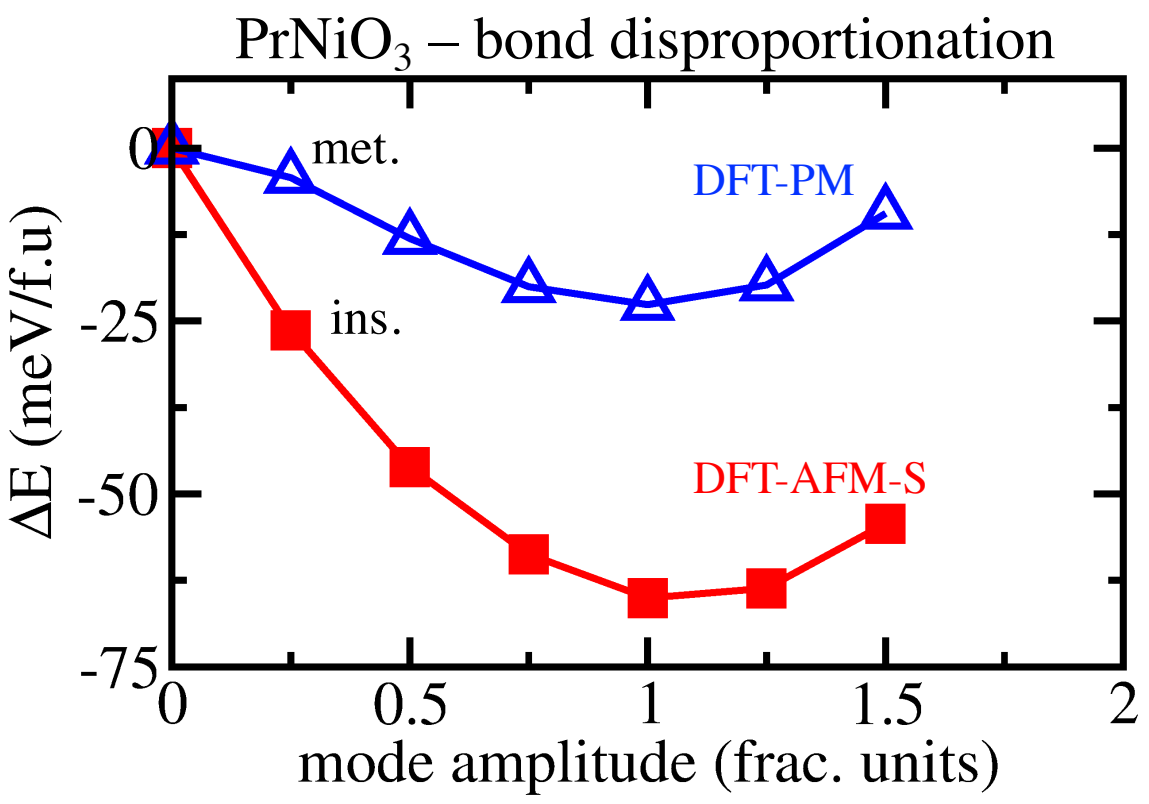

*Figure 6: Properties of isoelectronic compounds.* *Energy difference as a function of the bond disproportionation in $PrNiO_3$ in the AFM (squares) and PM (triangles) magnetic orders. Filled and unfilled symbols correspond to insulating and metallic solutions, respectively. The reference structure at 0 amplitude of JT or $B_{oc}$ modes is set to a material displaying octahedra rotations and anti-polar motions of ions.*

***The case of $PrNiO_3$*:** The $RNiO_3$ compounds form another important family of $ABO_3$ perovskites in which materials with an A site cation presenting a small ionic radius are insulating in both AFM and PM phases (R=Lu-Sm, Y) while remaining compounds (R=Nd, Pr) are only insulators in the AFM phase[72]. We correctly find that the geometry relaxation for $PrNiO_3$ with the AFM order results in an insulating monoclinic cell showing a band gap of 0.78 eV and disproportionation effects ($Q_{Boc}$=0.15 Å, $\mu_{NiL}$=1.38 $\mu_B$ and $\mu_{NiS}$=0 $\mu_B$), whose amplitudes are reduced with respect to $YNiO_3$. Just as in the case of $YNiO_3$, the stabilization of the breathing mode is not essential for the gap opening and the AFM order is already sufficient to produce a sizable band gap of 0.56 eV and an asymmetry of magnetic moments ($\mu_{NiL}$=1.21 $\mu_B$ and $\mu_{NiS}$=0 $\mu_B$) (see the energy potential surface as a function of $B_{oc}$ amplitude when starting from a cell with only rotations and anti-polar displacement presented in Fig6). Using the AFM structure but with a PM order, $PrNiO_3$ is also willing to adopt a disproportionated cell whose amplitude is a priori not enough to produce insulation (Fig6). In contrast to the AFM order and $YNiO_3$, the PM solution without bond disproportionation does not show any asymmetry of magnetic moments, all Ni cations bear a spin of 0.93 $\mu_B$. This observation signals the absence of a spontaneous electronic instability toward a disproportionation in the PM cell for $PrNiO_3$, in agreement with experiments. Nevertheless, a full structural relaxation with the PM order yields an insulating disproportionated cell at 0 K for $PrNiO_3$ with a narrow band gap of 0.30 eV in contrast with experiments – *i.e.* it becomes an insulator at the AFM transition ($T_N$=135 K [72]).

This discrepancy could result from an exaggerated electron localization in SCAN that contributes to (i) bond disproportionation and (ii) Ni magnetic moments substantially larger in SCAN-no-U ($Q_{Boc}$=0.17 and 0.15 Å, $\mu_{NiL}$=1.41 and 1.35 $\mu_B$ for R=Y and Pr, respectively) than in GGA+U calculations ($Q_{Boc}$=0.14 and 0.11 Å, $\mu_{NiL}$=1.26 and 1.17 $\mu_B$ for R=Y and Pr, respectively) of Refs.[20,24]. Overestimated magnetic moments are not specific to $ABO_3$ materials and instead seem inherent to the SCAN functional[27,73] that will have to be improved in this respect.

## V. <u>Conclusion</u>

We have shown that a DFT using a functional without on-site correlation energy U but better amending self-interaction errors captures the basic properties of 3d electron transition metal oxide perovskites namely band gap, magnetic moments, relative B-d to O-p levels positions and all structural features, by using a polymorphous representation allowing energy lowering formation of (i) magnetic moments, (ii) atomic displacements and octahedral rotations, and (iii) breaking of crystal field symmetry of partially occupied orbitals. We further show that lattice distortions including, Jahn-Teller and bond disproportionation modes, are captured by single-determinant, mean field DFT without U parameter suggesting that dynamic correlations are not the universal controlling factor here. Success of DFT for a given case does not exclude the existence of other forms of correlation. Furthermore, the success of DFT does not imply Coulomb interactions are not important, because DFT certainly includes (mean field like Hartree) Coulomb interaction. However, this is very different than the highly complex treatment of non-mean field Hubbard like U interaction term in the Hubbard Hamiltonian. The latter effect is excluded in the much simpler present calculation which still provides good and material-specific results.

***Acknowledgments***: This work received support from the European Research Council (ERC) Consolidator grant MINT under the Contract No. 615759. Calculations took advantages of the Occigen machines through the DARI project EPOC A0020910084 and of the DECI resource FIONN in Ireland at ICHEC through the PRACE project FiPSCO. The work of AZ was supported by Department of Energy, Office of Science, Basic Energy Science, MSE division under Grant No. DE-FG02-13ER46959 to CU Boulder. JV acknowledges technical support from

A. Ralph at ICHEC supercomputers.